\documentclass[a4paper,11pt]{article}
\pdfoutput=1 

\usepackage{jcappub} 

\usepackage[T1]{fontenc} 

\title{Probing the Purely Ingoing Nature of the Black-hole Event Horizon}

\author[a,b]{Adrian Ka-Wai Chung\note{Corresponding author.}, }
\author[a]{Tjonnie G.F.Li}

\affiliation[a]{Department of Physics, The Chinese University of Hong Kong, Shatin, N.T., Hong Kong}
\affiliation[b]{Theoretical Particle Physics and Cosmology Group, Department of Physics, King's College London, Strand, London, United Kingdom}

\emailAdd{ka-wai.chung@ligo.org}
\emailAdd{tgfli@cuhk.edu.hk}

\abstract{One of the most fundamental results of general relativity is that the event horizon of black hole is purely ingoing. 
On the other hand, semiclassical-gravity effects, such as particle creation and the quantization of black-hole area, suggest that black holes can emit energy.
Since a black hole is characterized by the presence of the event horizon, the emitted energy must be extracted from the black hole through its horizon. 
These considerations provide a motivation to test the validity of the purely ingoing nature of black-hole horizon. 
In this paper, we propose a novel test of the purely ingoing nature of black-hole horizon through gravitational-wave detection. 
We study the effects of hypothetical out-going gravitational waves to a perturbed black hole by supplementing the boundary condition of gravitational waves at the horizon with a phenomenological outgoing part.
We show that this leads to extra excitation of the usual quasinormal modes of a perturbed black hole and continuous emission of gravitational waves.  
These additional signatures enable us to test the boundary condition(s) of the black-hole event horizon through gravitational-wave detection. 
Reanalysing the merger remnant of GW150914, we constrain the intensity of the outgoing gravitational-horizon flux to be $ < 10^{40} \rm W$, which is roughly $ 10^{-9} $ of peak luminosity of GW150914.}

\begin{document}
\maketitle
\flushbottom

\section{Introduction}
\label{sec:intro}

The detection of gravitational waves emitted by black-hole mergers with the Advanced LIGO and Advanced Virgo detectors \cite{LIGO_detector_01, Virgo_detector_01} has opened a new window into the strong-field dynamics of gravity \citep{LIGO_08, LIGO_07, LIGO_03, LIGO_11, LIGO_12, LIGO_13}. 
In particular, by observing the settlement of a merger of a pair of black holes into a stationary one, a process known as the ringdown phase, we concluded that the spacetime of a remnant black hole is consistent with the description by the Kerr metric \cite{LIGO_11, Carullo_01, Carullo_02, Carullo_03, Brito_Buonanno_Raymond_01, Isi_01, NHT_TIGER_01, NHT_TIGER_02, NHT_TIGER_03, NHT_TIGER_04, IMR_Test_02}.
This conclusion in turns supports the existence of the event horizon of remnant black hole, where gravitational perturbations are purely ingoing \cite{Green_Function_Technique_01, Andersson_01}.  
This boundary condition determines the gravitational waves radiated in a discrete set of (complex) quasi-normal frequencies by a distorted remnant black hole during the ringdown phase (see e.g. \cite{Green_Function_Technique_02, QNM_03, QNM_04, QNM_05}). 

While the purely ingoing event horizon of black hole is one of the most fundamental results of general relativity, this boundary condition may not be the full story if one considers the quantum nature of black holes.
Semiclassical-gravity calculations suggest that particle creation \cite{BH_Radiation_01, BH_Radiation_02, BH_Radiation_03}, quantum fluctuation \cite{Quantum_fluctuation_01, Quantum_fluctuation_02, Quantum_fluctuation_03, Quantum_fluctuation_04}, tunneling \cite{Quantum_tunneling_01, Quantum_tunneling_02, Quantum_tunneling_03, Quantum_tunneling_04}, the polarization and decay of different possible vacuum states \cite{BH_vacca_01, BH_vacca_02, BH_vacca_03, BH_vacca_04, BH_vacca_05} and the quantization of black-hole area \cite{Area_Quantization_01, Area_Quantization_02, Area_Quantization_03, Area_Quantization_04, Area_Quantization_05, Area_Quantization_06, Area_Quantization_07, Area_Quantization_08, Area_Quantization_09, Area_Quantization_10, Area_Quantization_11} may happen at the event horizon of black holes at the quantum level. 
These processes lead to intrinsic radiation by black holes in the form of particles and waves. 
Therefore, at the quantum level, there could be outgoing waves, though extremely weak, at the event horizon due to these processes (see \cite{Book_01, Book_02, Book_03, Book_04, Book_05, Green_Function_Technique_02, Maggiore_vol_2} for reviews).
The aforementioned considerations constitute a motivated case to test the purely ingoing nature of the event horizon of astrophysical black holes.
The ringdown phase is an ideal probe of the boundary condition at the horizon because it involves deformation of the event horizon.
The ringdown signal is expected to be a useful probe of the nature of the remnant black hole.  
For example, by measuring the gravitational waves during the ringdown stage, we can accurately measure the mass and spin of the final black hole \cite{Carullo_02, QNM_04, QNM_05, Brito_Buonanno_Raymond_01, Nikhef_01, Nikhef_02, Ringdonw_01}. 
Moreover, the ringdown phase also offers us unprecedented opportunities to test aspects of general relativity, such as, to test the no-hair theorem \cite{Carullo_01, NHT_TIGER_01, NHT_TIGER_02, NHT_TIGER_03, NHT_TIGER_04, Isi_01}, measure the mass of the graviton mass \cite{Chung_01} and perform consistency tests of general relativity \cite{Carullo_03, BH_Nature_Test_01, IMR_Test_02, PhysRevLett.123.151101}. 

In this paper, we propose a novel method to test the purely ingoing nature of the black-hole event horizon by measuring hypothetical waves coming out of the event horizon of final black hole during its ringdown phase.
Throughout the paper a ``black hole with extend boundary conditions'' refers to a black hole that has an outgoing part in their boundary conditions representing hypothetical outgoing waves motivated the semiclassical effects mentioned above.  
Throughout the paper, we shall ignore effects to the final black hole by the outgoing waves, such as backreaction and the changes of absorption spectra \cite{BH_Scattering_01, BH_Scattering_02, BH_Scattering_03, BH_Scattering_04} due to the outgoing waves because the outgoing waves should be too small to change the overall macroscopic properties of black holes and just focus on the effects on gravitational-wave generation due to hypothetical outgoing waves. 
Unless otherwise specified, $ c = G = \hbar = 1$. 

\section{Possible boundary conditions at the event horizon of a black hole with extended boundary conditions}
\label{sec:BC}

The behaviour of a black hole in its ringdown phase can be studied by solving the Teukolsky equation. 
In the Boyer-Lindquist coordinates $ (t, r, \theta, \phi) $, where $ \theta $ is the angle between the line of sight and the spin of the black hole and $\phi$ is the azimuthal angle, the Teukolsky equation for vacuum perturbations is given by\cite{Teukolsky_01_PRL, Teukolsky_02_ApJ, Teukolsky_03_ApJ, Teukolsky_04_ApJ}
\begin{equation}
\begin{aligned}
& \left[\frac{\left(r^{2}+ M^2 a^{2}\right)^{2}}{\Delta}-M^2 a^{2} \sin ^{2} \theta\right] \frac{\partial^{2} \psi}{\partial t^{2}}+\frac{4 M^2 a r}{\Delta} \frac{\partial^{2} \psi}{\partial t \partial \phi} +\left[\frac{M^2 a^{2}}{\Delta}-\frac{1}{\sin ^{2} \theta}\right] \frac{\partial^{2} \psi}{\partial \phi^{2}} \\
& -\Delta^{-s} \frac{\partial}{\partial r}\left(\Delta^{s+1} \frac{\partial \psi}{\partial r}\right) -\frac{1}{\sin \theta} \frac{\partial}{\partial \theta}\left(\sin \theta \frac{\partial \psi}{\partial \theta}\right) -2 s\left[\frac{M a(r-M)}{\Delta}+\frac{i \cos \theta}{\sin ^{2} \theta}\right] \frac{\partial \psi}{\partial \phi} \\
& -2 s\left[\frac{M\left(r^{2}-M^2 a^{2}\right)}{\Delta}-r-i M a \cos \theta\right] \frac{\partial \psi}{\partial t} +\left(s^{2} \cot ^{2} \theta-s\right) \psi = 0, 
\end{aligned}
\end{equation}
where $ \Delta = (r - r_+)(r - r_-)$, $ r_+ $ and $ r_- $ are the outer and inner event horizon respectively, $ \psi = \rho^{-4} \psi_4 $ is the perturbation function for gravitational perturbations,  $ \psi_4 $ is the fourth Weyl scalar, $ \rho^{-1} = r - i M a \cos \theta $, $ M $ and $ a $ are the mass and the dimensionless spin of the black hole respectively, and  $ s $ is the spin weight of perturbation fields.
In particular, at spatial infinity ($ r \rightarrow \infty  $), $ \psi_4 \propto \ddot{h}_{+} - i \ddot{h}_{\times} $, where $ h_{+} $ and $ h_{\times} $ are the plus mode and cross mode polarization of gravitational waves. 
Since the Teukolsky equation is a separable differential equation, we let $ \psi(t, r, \theta, \phi) = R_{nlm} (r) S_{lm}(\theta, \phi) e^{- i \omega t}$, where $ S_{lm} (\theta, \phi) $ is the spheroidal function given $ l $ and $ m $\cite{Berti_01}. 

To thoroughly test the purely ingoing nature of black-hole  event horizon, we consider hypothetical gravitational waves coming out of the horizon and their effects on gravitational-wave generation during the ringdown phase. 
We represent the hypothetical outgoing waves by a phenomenological outgoing part at the boundary condition of gravitational perturbations of black hole,
\begin{equation}\label{eq:BC}
\begin{split}
R_{nlm}(x \rightarrow - \infty) = & R^{\rm in}_{nlm} (x) + R^{\rm out}_{nlm}(x), \\
R^{\rm in}_{nlm}(x) = & \mathcal{E}_{nlm}(\omega) \omega^2 \Delta^2 e^{- i k x }, \\
R_{nlm}^{\rm out} (x) = & \mathcal{H}_{nlm}(\omega) (\omega^{\dagger})^2 \Delta^2 e^{+i k^{\dagger} x + i \Phi_{\rm HE}}, 
\end{split}
\end{equation}
where $ x $ is the tortoise coordinate, $ k = \omega - m \Omega_H $, $ \Omega_H $ is the angular velocity of the event horizon, $ \mathcal{E}_{nlm}(\omega) $, $ \mathcal{H}_{nlm} (\omega) $ are the spectra of the ingoing and outgoing gravitational waves with angular dependence $ S_{lm} (\theta, \phi)$ at the horizon and $ \Phi_{\rm HE} $ is the phase of the outgoing part.

In Eq.~\eqref{eq:BC}, the outgoing waves are represented by the complex conjugate of the ingoing waves with corresponding frequencies, which can be argued for by the following reasons:
Firstly, in quantum mechanics, complex conjugation of wave functions corresponds to the time-reversal of quantum states\cite{Sakurai}.
This is consistent with our consideration that the emission of gravitational waves is the same as the time-reversal of their absorption\cite{Hawking_1976_QM}. 
Secondly, the complex conjugation ensures a time-decaying quasi-normal-mode response of a black hole with extended boundary conditions. 
Otherwise, this boundary condition may lead to solutions growing exponentially with time, which violates conservation of energy.  

\eqref{eq:BC} is different from the boundary conditions of gravitational-wave echoes \cite{Echoes_01, Echoes_02, Echoes_03, Echoes_04, Echoes_05, Echoes_06, Echoes_07, Echoes_08, Echoes_09, Echoes_10}, reflection of gravitational waves due to the microscopic structures of the event horizon suggested by quantum-gravity theories \cite{Information_paradox_01, Information_paradox_02, Information_paradox_03, Information_paradox_04, Information_paradox_05}. 
Firstly, in the echoes model, outgoing waves start propagating away from the black hole at some finite tortoise coordinate above the event horizon by reflection \cite{Echoes_03, Echoes_04, Echoes_05, Echoes_09}; 
whereas in our model, outgoing waves originate at the event horizon, at the limit of $ x \rightarrow + \infty $.
Secondly, in the echoes model, the energy spectrum of reflected waves are proportional to that of ingoing waves, following from the definition of reflection \cite{Echoes_04, Echoes_05, Echoes_Search_02, Echoes_Search_03}; 
whereas in our model, the energy spectra of the outgoing and ingoing waves are independent so that the former can take a completely different form.  
As we shall see in the coming section, these two unique features of our model result in long-lasting generation of gravitational waves, which may be detectable even though the ringdown phase is unobservable. 

We should also add here that we understand that this is not fully correct, but that this constitutes a phenomenological test. 
We recognize that \eqref{eq:BC} is a heuristic model which phenomenologically includes outgoing waves at the event horizon of a black hole.
Given that there are various candidates of quantum-gravity theories (such as \cite{string_theory, loop_quantum_gravity, Euclidean_quantum_gravity, Noncommutative_geometry, Supergravity}), the exact solution, properties and the generation of outgoing waves of such a black hole with the modified boundary conditions have yet to be fully explored. 
Nonetheless, as well shall see from the following investigation, \ref{eq:BC} can serve as our starting point for a phenomenological test. 

\section{Phenomenological description of the quasi-normal modes of a black hole with extended boundary conditions}
\label{sec:ringdown}

The ringdown of a black hole with extended boundary conditions can be studied by solving the Teukolsky equation subject to the emission boundary condition in Eq.~\ref{eq:BC}. 
By the Green's function technique\cite{Green_Function_Technique_01, Green_Function_Technique_02, Green_Function_Technique_03, Echoes_01, Echoes_02, Echoes_03, Echoes_04, Echoes_05, Echoes_06, Echoes_07, Echoes_08, Echoes_09, Echoes_10}, we obtain a gravitational waveform of the ringdown of a black hole with extended boundary conditions at the spatial infinity (see the Appendix for the detail of the derivation),
\begin{equation}\label{eq:QBH_RD_Waveform}
\begin{split}
h_{+} - i h_{\times} = & \frac{M}{r} \sum_{nlm} A_{nlm} \left( 1 +  C(\tilde{\omega}_{nlm})\right) S_{lm}(\theta, \phi) e^{- i \tilde{\omega}_{nlm} t} \\ 
&+ \frac{M}{r} \sum_{nlm} S_{lm}(\theta, \phi) \int_{0}^{+ \infty} \mathcal{H}_{nlm}(\omega) e^{- i \omega t} d \omega,
\end{split}
\end{equation}
where $ A_{nlm} $ are coefficients related to the quasi-normal modes of usual black holes.

From Eq.~\ref{eq:QBH_RD_Waveform}, we see that the usual quasi-normal-mode solution (i.e. term proportional to $A_{lmn}$) has been supplemented by an additional excitation factor, $ C(\tilde{\omega}_{nlm}) $, which are given by
\begin{equation}\label{eq:excitation_approx}
\begin{split}
C(\tilde{\omega}) \approx & \frac{\mathcal{H}(|\omega^{\rm Re}|) }{8 M^4 (\omega^{\rm Im})^{5}} \frac{(\tilde{\omega}^{\dagger})^2}{(\omega^{\rm Im})^2} \Big( \frac{4 M \omega^{\rm Im}}{e} \Big)^{4 M \omega^{\rm Im}} \\
& \times \Gamma(2 - 4 M \omega^{\rm Im}) \sum_{l = 0}^{2} (\omega^{\rm Im})^l \mathcal{Q}_l, \\
\end{split}
\end{equation}
where $ \mathcal{Q}_l $ are polynomials of $ \omega^{\rm Im} $ and $ \omega^{\rm Re}$, 
\begin{equation}
\begin{split}
\mathcal{Q}_0 = & 2 M\omega a^2 \Big[ C_1 \omega^{\rm Im} - 8(\omega^{\rm Im})^2 + a^2 \omega_{M}^{\rm Im} (\omega^{\rm Re})^2 \Big]\\
& + M \omega \Big[29 C_1 \omega^{\rm Im} + 8 C_2 (\omega^{\rm Im})^2 + 2 C_3 (\omega^{\rm Im})^3  \\
& \quad - 111 (\omega^{\rm Re})^2 + \omega_{M}^{\rm Im} \big(-3(2+a^2) C_1 \omega^{\rm Im} \\ 
& \quad - 2 a^2 C_2 (\omega^{\rm Im})^2 + 2(25 + 6a^2) (\omega^{\rm Re})^2 \big) \Big] \\
\mathcal{Q}_1 &= 15 C_1 + \omega^{\rm Im} \Big(6C_2+\omega^{\rm Im}(3C_3+2C_4\omega^{\rm Im}) \Big), \\
\mathcal{Q}_2 &= 45. \\
\end{split}
\end{equation}
The $C_i$ coefficients are given by
\begin{equation}
\begin{split}
C_1 &= 4 \omega^{\rm Re}, \\
C_2 &= \lambda^{\rm Im} + 2Mam\omega^{\rm Im} -12 M \omega^{\rm Re}, \\
C_3 &= 4 M a m - 2 \lambda^{\rm Im} M, \\
C_4 &= 16 M^3 a^2 \omega^{\rm Re} - 4 M^2 a m, 
\end{split}
\end{equation}
and $ \lambda^{\rm Im} $ is the separation constant of the Teukolsky equation. 
These terms are extra excitation of quasi-normal modes due to the emission of gravitational waves by the black-hole horizon. 

The terms proportional to $ \int_{0}^{+ \infty} \mathcal{H}_{nlm} (\omega) e^{- i \omega t} d \omega $ represent the continuous emission of gravitational waves by a black hole with extended boundary conditions. 
These terms correspond to an energy spectrum of gravitational waves perceived at spatial infinity of
\begin{align}\label{eq:energy_spectrum}
\frac{d E}{d \omega}& = \frac{\pi}{2} M^2 \omega^2 \sum_{nlm} |A_{nlm}|^2 |\mathcal{H}_{nlm} (\omega)|^2 \oint d \Omega |_{-2} S_{nlm}|^2 \nonumber \\
& \approx \frac{\pi}{2} M^2 \omega^2 \sum_{nlm} |A_{nlm}|^2 |\mathcal{H}_{nlm} (\omega)|^2, 
\end{align} 
where we have made use of the fact that $ \oint d \Omega |_{-2} S_{nlm}|^2 \approx 1 $\cite{Berti_01}. 

To implement our test for real detection of gravitational waves, we must assume a functional form for the spectrum of the outgoing waves, $ \mathcal{H}_{nlm}(\omega)$. 
In principle, if the outgoing waves have quantum-gravity (or semiclassical-gravity) origins, $ \mathcal{H}_{nlm}(\omega)$ can be calculated by considering the details of relevant quantum-gravity (semiclassical-gravity) processes. 
Nonetheless, few such theories have reached a state sufficient to have a fully consistent scheme for us to thoroughly explore the details of quantum-gravity processes which might generate outgoing waves (particles) at the horizon.
Despite these difficulties, we realize that the Bekenstein-Hawking entropy relation \cite{Bekenstein_01, BH_Radiation_01, BH_Radiation_02}, which implies the spectrum of Hawking radiation, is agreed by a number of proposed quantum-gravity theories (see, e.g., \cite{ST_Entropy_01, ST_Entropy_02, Loop_QG_Entropy_01, Loop_QG_Entropy_02}).
\textit{Inspired} by these considerations, in our work, we consider $ \mathcal{E} (\omega)$ of the form of the spectrum of Hawking radiation, as a \textit{heuristic} model, 
\begin{equation} \label{eq:graviton_emission_power}
\frac{d E}{d \omega} \propto \frac{\omega n(\omega)}{e^{\beta \omega} - 1}, 
\end{equation}
where $ n(\omega) $ is the greybody factor \cite{Book_02} and $ \beta = 1 /T_H $ is the reciprocal of the Hawking temperature. 
Since the Hawking temperature $ T_H \sim 10^{-8} \rm K $ for black holes of mass $ 5 M_{\odot} \leq M \leq 100 M_{\odot}$, the peak frequency of \ref{eq:graviton_emission_power} corresponds to $ \sim 10^3 \rm Hz $. 
Hence, to a good approximation, we can assume that most of the frequencies of the detector sensitivity band are smaller than the frequency corresponding to the peak of the emission spectrum. 
Therefore, we can take $ n(\omega) $ of the form of the low-frequency approximation \cite{Don_Page_01}, 
\begin{equation}
n(\omega) \approx M^6 \omega^6,  
\end{equation}
throughout the whole detection frequency band. 
Note that we have ignored the changes on $ n(\omega) $ and the absorption spectra due to the outgoing waves of the black holes, because we have assumed that the outgoing waves are small and cause no changes to macroscopic properties of the finally black holes. 
At last, we consider, $ \mathcal{H}_{nlm} (\omega) \propto \omega^2 (e^{\beta \omega} -1)^{-\frac{1}{2}} $. 
Since $ \sum_{nlm} |A_{nlm}|^2 \approx 1 $ \cite{Nikhef_01}, we have (with $ c, G $ and $ \hbar $ restored)
\begin{equation}\label{eq:H_omega}
\mathcal{H}_{nlm} (\omega) = A_{nlm} \left( \bar{\epsilon} \frac{L_{\rm GW}}{L (\beta)} \right)^{1/2} \sqrt{\frac{G \hbar}{c^5}} \Big( \frac{GM}{c^3} \Big)^2 \omega^2 (e^{\beta \omega} -1)^{-\frac{1}{2}}, 
\end{equation}
where we have parameterized the proportionality constant of \ref{eq:graviton_emission_power} in the form of $ \bar{\epsilon} L_{\rm GW}/L (\beta) $, where 
\begin{equation}
L(\beta) = \frac{d E}{d t} = \int_{0}^{+ \infty} d \omega \frac{M^6 \omega^7}{e^{\beta \omega} - 1}.  
\end{equation} 
Such parameterization enables us to compare the luminosity due to gravitational waves coming out of the horizon to the peak gravitational-wave luminosity $ L_{\rm GW}$ of the corresponding event.

$ \bar{\epsilon} $ characterizes the intensity of the hypothetical outgoing waves. 
If gravitational waves are purely ingoing at the black-hole horizon, then $ \bar{\epsilon} \equiv 0 $. 
If the measured $ \bar{\epsilon} > 0 $, the purely ingoing nature of the horizon of the detected remnant black hole is compromised. 
Thus measuring or constraining $ \bar{\epsilon}$ provides a test of the purely ingoing nature of the horizon of astrophysical black holes.

\begin{figure}[t!]
  \centering
	\includegraphics[width=\columnwidth]{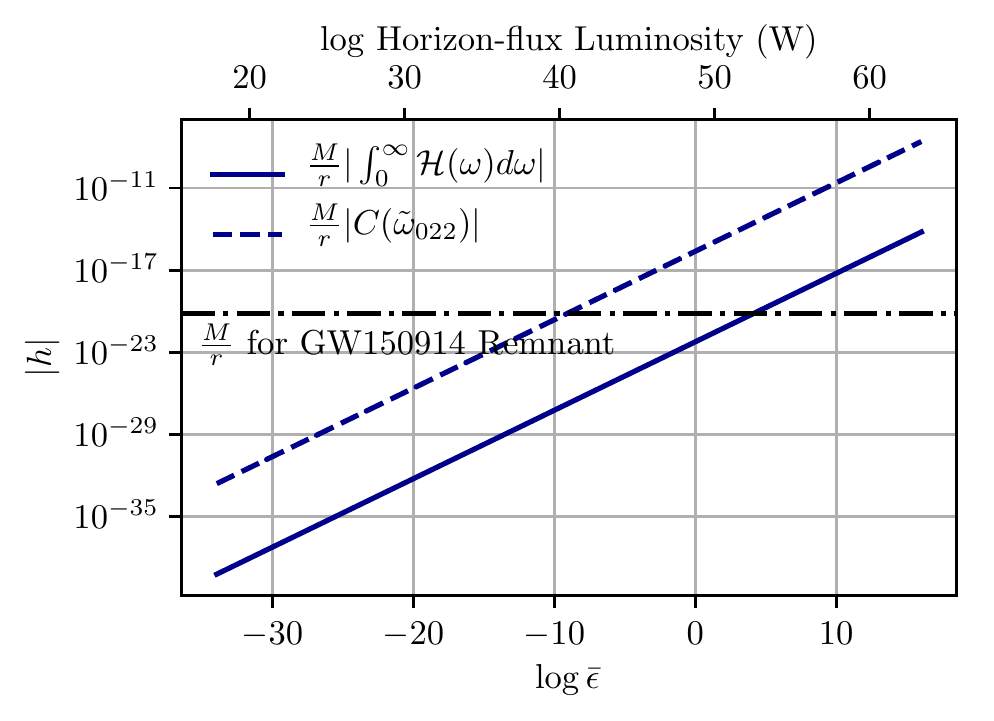}
	\caption{The magnitude of $ \frac{M}{r} |\int_{0}^{+ \infty} \mathcal{H}(\omega) d \omega| $ and $\frac{M}{r} |C(\tilde{\omega}_{022})|$ as a function of emissivity, with $ \tilde{\omega}_{022} $, $ M $ and $ r $ taken to be the estimated 022-mode frequency, final mass and luminosity distance of the GW150914 remnant. 
For references, the top axis plots the horizon-flux luminosity corresponds to the values of $ \bar{\epsilon} $ on the bottom axis. 
These two terms correspond to the continuous-emission term and extra quasinormal-mode excitation term in Eq.~\ref{eq:QBH_RD_Waveform} at the start of the ringdown phase. 
The horizontal dashed line plots the value of $ \frac{M}{r} $ of the GW150914 remnant. 
From the plot, we conclude that $ |C(\tilde{\omega}_{022})| $ is much larger than $ |\int_{0}^{+ \infty} \mathcal{H}(\omega) d \omega| $. 
Thus, the extra excitation of quasi-normal modes is comparatively more visible than the continuous emission.
}
  \label{fig:Term_Contribution}
\end{figure}

Fig.~\ref{fig:Term_Contribution} plots the magnitude of $ \frac{M}{r} |\int_{0}^{+ \infty} \mathcal{H}(\omega) d \omega| $ and $\frac{M}{r} |C(\tilde{\omega}_{022})|$ as a function of $ \bar{\epsilon} \in [10^{-35}, 10^{18}] $, with $ \tilde{\omega}_{022} $, $ M $, $ r $ and $ \beta $ taken to be the estimated 022-mode frequency, final mass, luminosity distance and the reciprocal Hawking temperature of the GW150914 remnant.
The top axis plots the luminosity of outgoing gravitational-horizon flux corresponding to the values of $ \bar{\epsilon} $ on the bottom axis. 
The plot gives an estimation of the orders of magnitude of the continuous-emission and  extra quasi-normal-mode excitation term in Eq.~\ref{eq:QBH_RD_Waveform} at the start of the ringdown phase. 
For $ \bar{\epsilon} \in [10^{-35}, 10^{18}]$, we see that $ |\int_{0}^{+ \infty} \mathcal{H}(\omega) d \omega| $ and $ |C(\tilde{\omega}_{022})| $ are both increasing with $ \bar{\epsilon}^{1/2} $. 
In general, $ |C(\tilde{\omega}_{022})| $, additional excitation of quasinormal modes, is larger than $ | \int_{0}^{+ \infty} \mathcal{H}(\omega) d \omega | $, the continuous emission of the horizon flux. 
Since $ |\tilde{\omega}| $ of other overtones are of similar order as the dominant mode, we expect similar order of $ |C(\tilde{\omega})| \sim |C(\tilde{\omega}_{022})| \gg |\int_{0}^{+ \infty} \mathcal{H}(\omega) d \omega| $ for other overtones in the ringdown phase. 
From Fig.~\ref{fig:Term_Contribution}, we conclude that the extra excitation of  black-hole quasinormal-mode responses should dominate over the other contributions for $ \bar{\epsilon} > 0 $.
In particular, as $ \bar{\epsilon} $ approaches $ \sim 10^{-10} $, the effects of additional excitation of quasinormal modes becomes as visible as the usual ringdown gravitational waves for GW150914. 

\section{Results of GW150914 Remnant} 
\label{sec:gw150914}

We use the waveform in Eq.~\ref{eq:QBH_RD_Waveform} with $ \mathcal{H}_{nlm} (\omega) $ given by Eq.~\ref{eq:H_omega} to re-analyse 4096 seconds of data around GW150914 \cite{GWOSC_01}.  
The signal is band-passed in the band [20,2038] Hz interval. 
In particular, we estimate the parameters associated with the ringdown of the remnant of GW150914, with the base-10 log of the emissivity, $ \log \bar{\epsilon} $ and $ \Phi_{\rm HE}$, used as free parameters.
According to Bayes's theorem, the posterior is proportional to the product of the likelihood and the prior
\begin{equation}\label{eq:Bayes_theorem}
p(\log \bar{\epsilon}, \Phi_{\rm HE}|d, H, I) \propto p(d| \log \bar{\epsilon}, \Phi_{\rm HE}, H, I) p(\log \bar{\epsilon}, \Phi_{\rm HE}|H), 
\end{equation}
where $ p(\log \bar{\epsilon}, \Phi_{\rm HE}|H)$ is the prior of $ \bar{\epsilon} $ and $ \Phi_{\rm HE} $ and $p(d| \log \bar{\epsilon}, \Phi_{\rm HE}, H, I) $ is the likelihood of a black hole with extended boundary conditions of emissivity $ \log \bar{\epsilon} $ producing a signal $ d $. 
The prior uniform of $ \bar{\epsilon} $ has been choosen to be uniform between $ \log \bar{\epsilon} \in [-35, 15]$.
This range of prior covers the peak luminosity of gravitational radiation of GW150914 so that we can thoroughly test the possibility of unusually strong gravitational waves coming out of the horizon.  
The natural log of the likelihood is given by \cite{Carullo_02}
\begin{equation}
\begin{split}
& \ln p(d | \log \bar{\epsilon}, H, I) =-\frac{1}{2} \iint d t d \tau (d(t)-h(t|\bar{\epsilon})) \mathcal{C}^{-1}(\tau)(d(t+\tau)-h(t+\tau|\bar{\epsilon}))
\end{split}
\end{equation}
where $ d $ denotes the detected signals, $ h(t|\bar{\epsilon}) $ denotes the template waveform (i.e. Eq.~\ref{eq:QBH_RD_Waveform}) and $ \mathcal{C}(\tau) $ is the two-point autocovariance function of noise, defined by
\begin{equation}
\mathcal{C}(\tau) = \int dt n(t) n(t + \tau), 
\end{equation}
where $ n(t) $ denotes the noises. 
To ensure that there is no contamination from the merger into ringdown signal, we choose the lower bound of the prior of the ringdown start time to be $ 15 M $ after the merger\cite{Carullo_01}, with $ M \sim 68 M_{\odot} $ is the estimated final mass of GW150914 remnant \cite{LIGO_01, LIGO_07}. 
For this work, we only include the $nl|m|=022$ modes, because these modes have been found to be dominant \cite{Nikhef_01}, and the sole inclusion of these are sufficient for accurate ringdown spectroscopy \cite{Carullo_02}. 
Finally, we make use of the \texttt{pyRing} package, which is introduced to perform the analysis presented in \cite{Carullo_02}, to sample the posterior by the Bayesian nested-sampling algorithm \texttt{CPNest} \cite{CPNest}.

\begin{figure}[t!]
  \centering
	\includegraphics[width=\columnwidth]{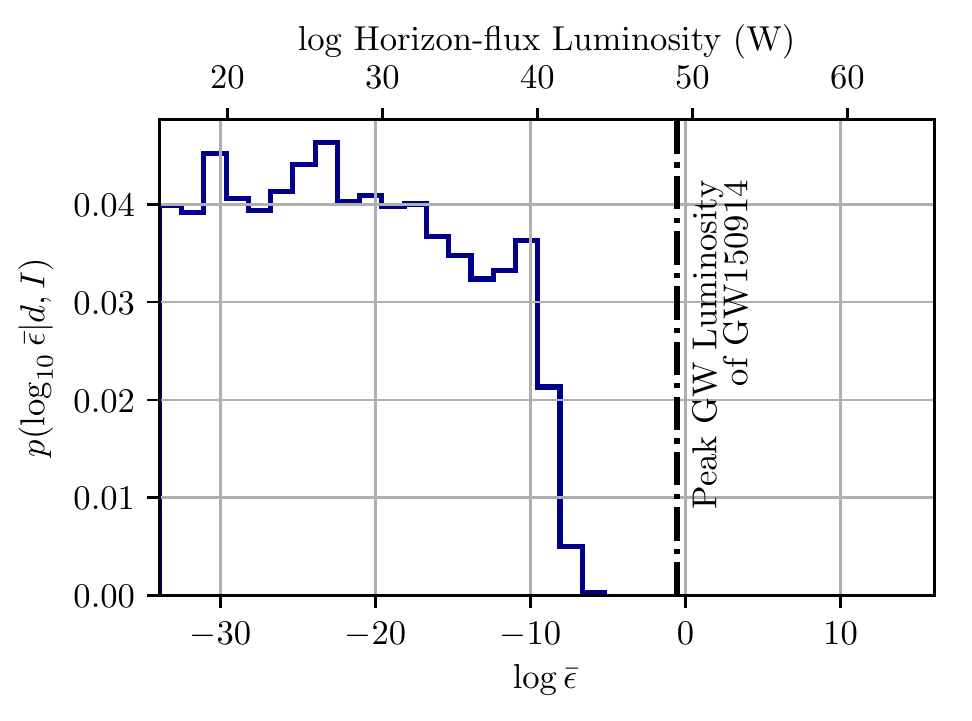}
	\caption{The posterior of $\bar{\epsilon}$ of the remnant black hole of GW150914, $p(\log \bar{\epsilon} |d, H, I) $. 
For references, the vertical dashed line plots the peak gravitational-radiation luminosity of GW150914. 
The posterior decreases sharply as $ \log \bar{\epsilon} $ increases from zero and shows no support for horizon-flux luminosity $ \lessapprox 10^{40} \rm W$, corresponding to $ \bar{\epsilon} \approx 10^{-10}$, suggesting that we find no evidence of outgoing gravitational-horizon flux detected. 
}
\label{fig:Summary_Plot}
\end{figure}

Fig.~\ref{fig:Summary_Plot} shows the posterior of $ \log \bar{\epsilon} $ for the final black hole of GW150914, marginalized over other parameters, such as mass and spin, and phase $ \Phi_{\rm HE}$.  
We have checked that the posterior of $ \log \bar{\epsilon} $ is not degenerate with the posterior of other parameters of the final black hole of GW150914.
The posterior is consistent with $ \bar{\epsilon} = 0$ and shows no significant support for $ \bar{\epsilon} > 10^{-10}\ $, which sets the constraint on the luminosity of outgoing gravitational flux at the event horizon to be $ \lessapprox 10^{40} \rm W$ for the final black hole of GW150914.
As our constraint is extremely small compared to the peak gravitational-wave luminosity of GW150914, we conclude that we have found no evidence of outgoing gravitational-horizon flux of the remnant of GW150914, being detected. 

\section{Concluding Remarks} 

Motivated by the semiclassical-gravity effects of black holes \cite{BH_Radiation_01, BH_Radiation_02, BH_Radiation_03, Quantum_fluctuation_01, Quantum_fluctuation_02, Quantum_fluctuation_03, Quantum_fluctuation_04, Quantum_tunneling_01, Quantum_tunneling_02, Quantum_tunneling_03, Quantum_tunneling_04, BH_vacca_01, BH_vacca_02, BH_vacca_03, BH_vacca_04, BH_vacca_05, Area_Quantization_01, Area_Quantization_02, Area_Quantization_03, Area_Quantization_04, Area_Quantization_05, Area_Quantization_06, Area_Quantization_07, Area_Quantization_08, Area_Quantization_09, Area_Quantization_10, Area_Quantization_11}, we propose a novel test to probe the purely ingoing nature of the black-hole event horizon by measuring hypothetical waves coming out of the event horizon of final black hole during its ringdown phase.
We supplement the purely ingoing boundary condition with an phenomenological outgoing part. 
The outgoing part adds an outgoing gravitational-horizon flux to the gravitational waves by a remnant black hole during its ringdown phase and thus opens up the possibility of testing the purely ingoing nature of black-hole horizon through detecting the ringdown phase. 
Reanalysing the remnant of GW150914, we constrain its outgoing gravitational-horizon flux to be $ \lessapprox 10^{40} \rm W $, corresponding to about $ 10^{-9} $ of the peak gravitational-wave luminosity of GW150914.
Our stringent results suggest that we have found no evidence of outgoing gravitational horizon flux coming out from an astrophysical black holes. 
Along with searches of echoes of gravitational waves \cite{Echoes_01, Echoes_02, Echoes_03, Echoes_04, Echoes_05, Echoes_06, Echoes_07, Echoes_08, Echoes_09, Echoes_10}, our test helps providing a more thorough understanding of the properties of black holes.

Throughout this work, we have made a number of simplifications. 
Firstly, we neglect possible effects of black-hole supperadiance \cite{BH_Superradiance, BH_Superradiance_02} and the Penrose process \cite{Penrose_Process}, which might also generate outgoing gravitational waves at the horizon. 
This is acceptable as long as only the upper limit on the outgoing waves at the horizon is concerned. 
With this simplification, we can perform a maximum-reach analysis by attributing all outgoing radiation, if any, as the effects of the outgoing waves hypothesized in this work. 
Secondly, we also ignore the changes of absorption spectra \cite{BH_Scattering_01, BH_Scattering_02, BH_Scattering_03, BH_Scattering_04} something more concrete about the changes induced by included the absorption spectra, which is related to the ratio between the incident waves and scattered waves as observed at spatial infinity. 
The outgoing waves will change the content of scattered waves which in turn alter the absorption cross section.
In principle, we should also include these effects into our considerations for our test. 
Nonetheless, these effects can be neglected if the outgoing waves are small, as our constraint suggests. 
This aspect would be studied and included for more accurate and improved constraints in the future. 

We expect that the constraints of our test will be improved progressively in the future. 
In this analysis for GW150914, due to the limited detector sensitivity sensitivity, we have only included the dominant quadrupole mode ($nl|m|=022$) for our constraints.
With the improvement of sensitivity of the existing ground based gravitational-wave detections, along with the launch of the next-generation detectors, for example \cite{KAGRA, Einstein_Telescope, InDIGO, Cosmic_Explorer}, higher modes of the ringdown phase can be measured with increasing precision \cite{LIGO_12, LIGO_13}. 
By including more higher modes into our waveform models (i.e. more terms in \ref{eq:QBH_RD_Waveform}), more contrasting features are included for us to pose even more stringent constraints on the out-going gravitational horizon flux.
Along with other existing searches \cite{Echoes_Search_01, Echoes_Search_02, Echoes_Search_03, Echoes_Search_04, Echoes_Search_05, Echoes_Search_06, Echoes_Search_07, Echoes_Search_08}, our test helps offering a more complete understanding of the nature of the event horizon of black hole. 

\appendix
\section{Derivation of the Ringdown Waveform of a black hole with extended boundary conditions} 
\label{Appendix A}

The Teukolsky equation is a separable linear second order differential equation \cite{Teukolsky_01_PRL, Teukolsky_02_ApJ, Teukolsky_03_ApJ, Teukolsky_04_ApJ}, For gravitational perturbations, we let $ u (r, t) = \sqrt{r^2+M^2 a^2} \Delta^{-1} R (r, t) $. 
The radial Teukolsky equation in the time-domain becomes, 
\begin{equation}
\frac{\partial^2 u}{\partial x^2} - \frac{\partial^2 u}{\partial t^2} - V(r) u = 0 
\end{equation}
where $ x $ is the tortoise coordinate defined by $ \frac{d}{d x} = \frac{\Delta}{r^2+a^2} \frac{d}{dr} $ and $ V(r) $ is the complex effective potential. 
If we consider the Laplace transform of $ u(r, t) $,
\begin{equation}\label{eq:Laplace_Transform}
\tilde{u}(r, \omega) = \int_{0}^{+ \infty} u(r, t) e^{i \omega t} dt, 
\end{equation}
then $ u(r, \omega) $ satisfies the equation of 
\begin{equation}\label{eq:SE}
\frac{\partial^2 \tilde{u}}{\partial x^2} + \Big( \omega^2 - V(r) \Big) \tilde{u} = \mathcal{I}(x, \omega), 
\end{equation}
where $ \mathcal{I} = i \omega u_0 - \partial_t u_0 $ is the initial data of the quasi-normal-mode excitation, $ \partial_t u_0 = \partial_t u(t=0, x) $ and $ u_0 = u(t=0, x) $.
When $ \mathcal{I} = 0 $, Eq.~\ref{eq:SE} admits two linearly independent solutions: the down mode, 
\begin{equation}\label{eq:down_mode}
\tilde{u}_{\rm down} (x, \omega) \approx 
\begin{cases}
& \Delta e^{- i k x}, ~ x \rightarrow r_+ \\
& A_1 (\omega) r^{-2} e^{- i \omega r} + A_2(\omega) r^2 e^{i \omega r} , ~ x \rightarrow + \infty \\ 
\end{cases}
\end{equation}
which is purely ingoing at the event horizon, and the up mode
\begin{equation}\label{eq:up_mode}
\tilde{u}_{\rm up} (x, \omega) \approx 
\begin{cases}
& B_1(\omega) \Delta e^{-i k x} + B_2 (\omega) \Delta^{-1} e^{i k x}, ~ x \rightarrow r_+ \\
& r^{2} e^{i \omega r} , ~ x \rightarrow + \infty \\ 
\end{cases}
\end{equation}
which is purely outgoing at spatial infinity, where $ A_1(\omega) $, $A_2(\omega) $, $ B_1 (\omega) $ and $ B_2 (\omega)$ are functions of $ \omega$. 
In particular, $ A_1(\omega) $ and $ B_2 (\omega)$ are proportional to the Wronskian of the differential equation.

When $ \mathcal{I} \neq 0 $, Eq.~\ref{eq:SE} can be solved by the Green's function technique.
The desired Green's function can be constructed by Wronskian $ W(\omega) $, a function of frequency, of Eq.~\ref{eq:down_mode} and Eq. \ref{eq:up_mode}. 
By definition, quasi-normal-mode frequencies $\tilde{\omega}_{nlm} $ are complex zeroes of $ W(\omega) $ (for this reason, $ A_1 (\omega) \propto B_2 (\omega) \propto W(\omega) $ contribute no amplitude to the ringdown waveform). 
Thus, for a given set of $l$ and $ m$,  
\begin{equation}\label{eq:Wronskian}
\frac{1}{W(\omega)} \approx \frac{1}{2 \pi i} \frac{1}{ M^4 } \sum_{n} \frac{\zeta_{nlm}}{\omega - \tilde{\omega}_{nlm}}, 
\end{equation}
where $M^4$ is a factor for dimensional consistency and $ \zeta_{nlm} $ are constants which can be determined by the classical quasi-normal-mode solutions at spatial infinity (corresponding to the given $ l, m$), 
\begin{equation}\label{eq:NR_result}
\tilde{u} (x \rightarrow + \infty, t) \approx M r^2 \sum_{n} (\tilde{\omega}_{nlm})^2 A_{nlm} e^{- i \tilde{\omega}_{nlm} t}, 
\end{equation} 
where $ A_{nlm} = \sum_j a_j e^{i \delta_j} \eta^j $ are function of mass, symmetric mass ratio and spins of the parental black holes \cite{Nikhef_01, Nikhef_02}. 
If we choose 
\begin{equation}
\mathcal{I}(x, \omega) = \lim_{x_0 \rightarrow - \infty} M \omega^2 \Delta^{-1} e^{i k x} \delta(x - x_0), 
\end{equation}
and 
\begin{equation}
\zeta_{nlm} = A_{nlm}, 
\end{equation}
then upon performing the inverse Laplace transform, 
\begin{equation}\label{eq:inverse_laplace_transform}
u(x, t) = \frac{1}{2 \pi i} \lim_{\epsilon \rightarrow 0} \int_{-\infty + i \epsilon}^{\infty + i \epsilon} d \omega \tilde{u}(x, \omega) e^{- i \omega t}, 
\end{equation}
with the integration contour choosen to be closed by a semicircle centred at $ \omega^{\rm Im} = 0 $ covering negative imaginary axis \cite{Green_Function_Technique_01, Green_Function_Technique_02}, the form of Wronksian Eq.~\ref{eq:Wronskian} recovers the ringdown waveform of classical black holes.

Note that the outgoing wave $ \Delta e^{+i k^{\dagger} x} $ does not satisfy the radial Teukolsky equation Eq.~\ref{eq:SE}. 
Instead, it satisfies the complex conjugate of the equation.
To simplify the calculations, we construct an auxiliary function $ z = \tilde{u} - (\omega^{\dagger})^2 \mathcal{H}(\omega) \Delta e^{+i k^{\dagger} x} $, which satisfies 
\begin{equation}\label{eq:IHSE}
\frac{\partial^2 z}{\partial x^2} + \Big( \omega^2 - V(r) \Big) z = 2 i \Big(V_{\rm Im}(r) - \omega_{\rm Im}^2 \Big) \mathcal{H}(\omega)(\omega^{\dagger})^2 \Delta e^{+i k^{\dagger} x},
\end{equation}
where $ V_{\rm Im}(r) $ is the imaginary part of the effective potential, 
\begin{equation}
\begin{split}
V_{\rm Im} (r) \approx & \frac{4 \omega^{\rm Re} r + \lambda^{\rm Im} + 2M a \omega^{\rm Im} m - 12 M \omega^{\rm Re}}{r^2 + M^2 a^2} \\
& + \frac{(4 M a m - 2 \lambda^{\rm Im} M)r	+ 16 M^3 a^2 \omega^{\rm Re} - 4 M^2 a m}{(r^2+M^2 a^2)^2}, \\ 
\end{split}
\end{equation}
where $ \omega^{\rm Re} $ and $ \omega^{\rm Im} $ are respectively the real part and imaginary part of $ \omega $, $ m $ is the azimuthal number of the quasi-normal mode, $ a $ is the dimensionless spin of the black hole and $ \lambda^{\rm Im} $ is the imaginary part of the separation constant $ \lambda $.

The boundary condition of Eq.~\ref{eq:IHSE} is purely ingoing at the event horizon: $ z (x \rightarrow - \infty ) \propto \Delta e^{-i k x} $. 
Thus, its particular solution can be obtained by applying the Green's function technique. 
Using the Green's function technique, we have 
\begin{equation}
\begin{split}\label{eq:z_w}
z(x, \omega) = 2i \tilde{u}_{\rm up}(x, \omega) \frac{\mathcal{H}(\omega) C(\omega)}{W(\omega)}, 
\end{split}
\end{equation}
where $ C(\omega) $ is the excitation factor given by 
\begin{equation}\label{eq:excitation factor}
\begin{split}
C(\omega) \approx i \int_{-\infty}^{+ \infty} dx' \tilde{u}_{\rm down} (x') \Big(V_{\rm Im}(r') - \omega_{\rm Im}^2 \Big) (\omega^{\dagger})^2 \Delta' e^{+i k^{\dagger} x'}, 
\end{split}
\end{equation} 
where $ r' = r(x') $, $ \Delta' = (r' - r_+)(r' - r_-) $.
The above integral is approximately given by Eq.~(4) in the main text.

We then perform the inverse Laplace transform to Eq.~\ref{eq:z_w}, yielding
\begin{equation}\label{eq:z_t}
\begin{split}
z(r \rightarrow + \infty, t) = M r^2 \sum_{nlm} & (\tilde{\omega}_{nlm}^{\dagger})^2 A_{nlm} \mathcal{H}(\tilde{\omega}_{nlm}) C (\tilde{\omega}_{nlm}) r^2 e^{- i \tilde{\omega}_{nlm} t}. 
\end{split}
\end{equation}
If we redefine $ C(\tilde{\omega}) = \mathcal{H}(\tilde{\omega}) C (\tilde{\omega})$, then the time-domain ringdown waveform Eq.~(3) in the main text follows from the above calculations.

The above calculations can similarly be done for different $ l $ and $ m $.  
Adding the contribution from all quasi-normal modes at the spatial infinity where we detect gravitational waves, we obtain the waveform in Eq.~(3) in the main text. 

\acknowledgments

The authors are grateful to Gregorio Carullo and Walter Del Pozzo for their indispensable assistance in using \texttt{pyRing}. 
The authors would like to thank Pak-Tik Fong, Pak-Ming Hui, Gideon Koekoek, Feng-Li Lin and Lezhi Lo for the in-depth discussion in quantum mechanics, Giulia Pagano and Issac C.F. Wong for their assistance in programming, and Chris Van Den Broeck, Mark H.Y. Cheung, Joseph Gais, Louis Hamaide, Alvin K.Y. Li, Alan T.L. Lam, Harald Pfeiffer and Mairi Sakellariadou for their comments on this manuscript. 
A.K.W.C was supported by the Hong Kong Scholarship for Excellence Scheme (HKSES). 
The work described in this paper was partially supported by grants from the Croucher Foundation of Hong Kong, the Research Grants Council of Hong Kong (Project No. CUHK14306218) and the Research Committee of the Chinese University of Hong Kong.

This research has made use of data, software and/or web tools obtained from the Gravitational Wave Open Science Center (https://www.gw-openscience.org), a service of LIGO Laboratory, the LIGO Scientific Collaboration and the Virgo Collaboration. LIGO is funded by the U.S. National Science Foundation. Virgo is funded by the French Centre National de Recherche Scientifique (CNRS), the Italian Istituto Nazionale della Fisica Nucleare (INFN) and the Dutch Nikhef, with contributions by Polish and Hungarian institutes.
This paper carries a LIGO document numbers of P2000068-v2 and a document number of KCL-PH-TH/2019-99. 


\bibliographystyle{plain}
\bibliography{bibtex}



\end{document}